\documentclass[12pt]{article}
\pdfoutput=1   
\usepackage[margin=1.0in]{geometry}
\usepackage{amsmath}
\usepackage{amssymb}
\usepackage{amsfonts}
\usepackage{parskip}
\usepackage{natbib}
\usepackage[margin=1cm]{caption}
\usepackage{adjustbox}
\usepackage[toc,page]{appendix}
\usepackage{enumitem}
\usepackage{multirow}
\usepackage{setspace}
\usepackage{url}
\usepackage{xr}

     {\begin{list}{}%
             {\setlength{\leftmargin}{#1}}%
             \item[]%
     }
     {\end{list}}

\newcommand{\vn}{\textbf{n}}
\newcommand{\vp}{\textbf{p}}
\newcommand{\vX}{\textbf{X}}
\newcommand{\vZ}{\textbf{Z}}
\newcommand{\vbeta}{\boldsymbol{\beta}}
\newcommand{\vxi}{\boldsymbol{\xi}}

\newcommand{\Exp}{\mbox{Exp}}
\newcommand{\Ga}{\mbox{Ga}}
\newcommand{\We}{\mbox{We}}
\newcommand{\LN}{\mbox{LN}}
\newcommand{\Po}{\mbox{Po}}
\newcommand{\Mult}{\mbox{Mult}}

\newcommand{\pdet}{p^{(det)}}
\newcommand{\ind}{\stackrel{ind}{\sim}}
\newcommand{\Fm}{F_T^{(C)}}
\newcommand{\dt}{\text{det}}
\newcommand{\ftdt}{f_{T|\dt}(t|\dt)}
\newcommand{\nobs}{n^{(obs)}}

\providecommand{\blind}{0}

\title{Assessing the impacts of time to detection distribution assumptions on detection probability estimation}

\if0\blind
\author{Adam Martin-Schwarze, Jarad Niemi, and Philip Dixon\\
Department of Statistics, Iowa State University, Ames, Iowa}
\fi

\begin{document}

\maketitle
\newpage

\begin{abstract}

Abundance estimates from animal point-count surveys require accurate estimates of detection probabilities.  
The standard model for estimating detection from removal-sampled point-count surveys assumes that organisms at a survey site are detected at a constant rate; however, this assumption can often lead to biased estimates.  
We consider a class of N-mixture models that allows for detection heterogeneity over time through a flexibly defined time-to-detection distribution (TTDD) and allows for fixed and random effects for both abundance and detection.
Our model is thus a combination of survival time-to-event analysis with unknown-N, unknown-p abundance estimation.  
We specifically explore two-parameter families of TTDDs, e.g. gamma, that can additionally include a mixture component to model increased probability of detection in the initial observation period.
Based on simulation analyses, we find that modeling a TTDD by using a two-parameter family is necessary when data have a chance of arising from a distribution of this nature.
In addition, models with a mixture component can outperform non-mixture models even when the truth is non-mixture.  
Finally, we analyze an Overbird data set from the Chippewa National Forest using mixed effect models for both abundance and detection.
We demonstrate that the effects of explanatory variables on abundance and detection are consistent across mixture TTDDs but that flexible TTDDs result in lower estimated probabilities of detection and therefore higher estimates of abundance. 


\end{abstract}

\newpage
\section{Introduction}\label{sec:intro}

Abundance estimates from animal point-count surveys require accurate estimates of detection probabilities.  
Removal sampling, where individuals are solely counted on their first capture, provides one established methodology for estimating detection probabilities \citep{Farnsworth2002}.
A standard assumption in removal sampling is a constant detection rate throughout the observation period, but this assumption is often unjustified \citep{Alldredge2007, LeeMarsden2008}.  
In particular, animal behaviors such as intermittent singing in birds and frogs or diving in whales \citep{Scott2005, Diefenbach2007, Reidy2011}, differences in behavior across subgroups of animals \citep{Otis1978, Farnsworth2002}, observer impacts on animal behaviors \citep{McSheaRappole1997, Rosenstock2002, Alldredge2007}, and variations in observer effort, e.g. saturation or lack of settling down period \citep{Petit1995, LeeMarsden2008, Johnson2008}, can all lead to time-varying rates of detection.

In this manuscript, we develop a model for scenarios where detection rates are not constant over time. 
We analyze times to first detection as time-to-event data, as is done in parametric survival analysis, defining a continuous random variable $T$ for each individual's time to first detection with a probability density function (pdf) $f_T(t)$ and cumulative distribution function (cdf) $F_T(t)$.  
We refer to the distribution of $T$ as a time-to-detection distribution (TTDD).

One common strategy to deal with data that do not fit a constant-detection assumption is to model the TTDD as a mixture of two distributions --- a continuous-time distribution and a point mass for increased detection probability in the initial observation period \citep{Farnsworth2002, Farnsworth2005, EffordDawson2009, Etterson2009, Reidy2011}.
However, this is not yet the standard \citep{Solymos2013, Amundson2014, Reidy2016}. 
We consider the choice of whether to include a mixture component in conjunction with TTDDs with non-constant rates
and apply the term \emph{mixture TTDD} when the TTDD has a discrete and continuous component.

Unlike most survival analyses, the number of individuals $N$ present at a survey is unknown and may be the primary quantity of interest.  
We embed the TTDD in a hierarchical framework for multinomial counts called an N-mixture model \citep{Wyatt2002, Royle2004NMixture}, which is an entirely different use of `mixture' from the mixture models in the previous paragraph. 
For our purposes, the N-mixture framework provides three clear benefits: 1) it handles counts within a flexible multinomial data framework \citep{RoyleDorazio2006} which accords with the interval-censored data collection that is customary in point-count surveys \citep{Ralph1995}, 2) the hierarchical structure readily lends itself to including abundance- and detection-related covariates and random effects \citep{Dorazio2005, Etterson2009, Amundson2014}, and 3) for a Bayesian analysis, we can sample the posterior joint distribution of N-mixture parameters straight-forwardly using Markov chain Monte Carlo (MCMC).  
The N-mixture framework models abundance as a latent variable with a Poisson or other discrete distribution and independently models detection probabilities.  
Several previous studies have employed the N-mixture framework to analyze removal sampled point-count data  \citep{Royle2004Generalized, Dorazio2005, Etterson2009, Solymos2013, Amundson2014, Reidy2016}.  

\begin{figure}\centering
\includegraphics[width=\textwidth]{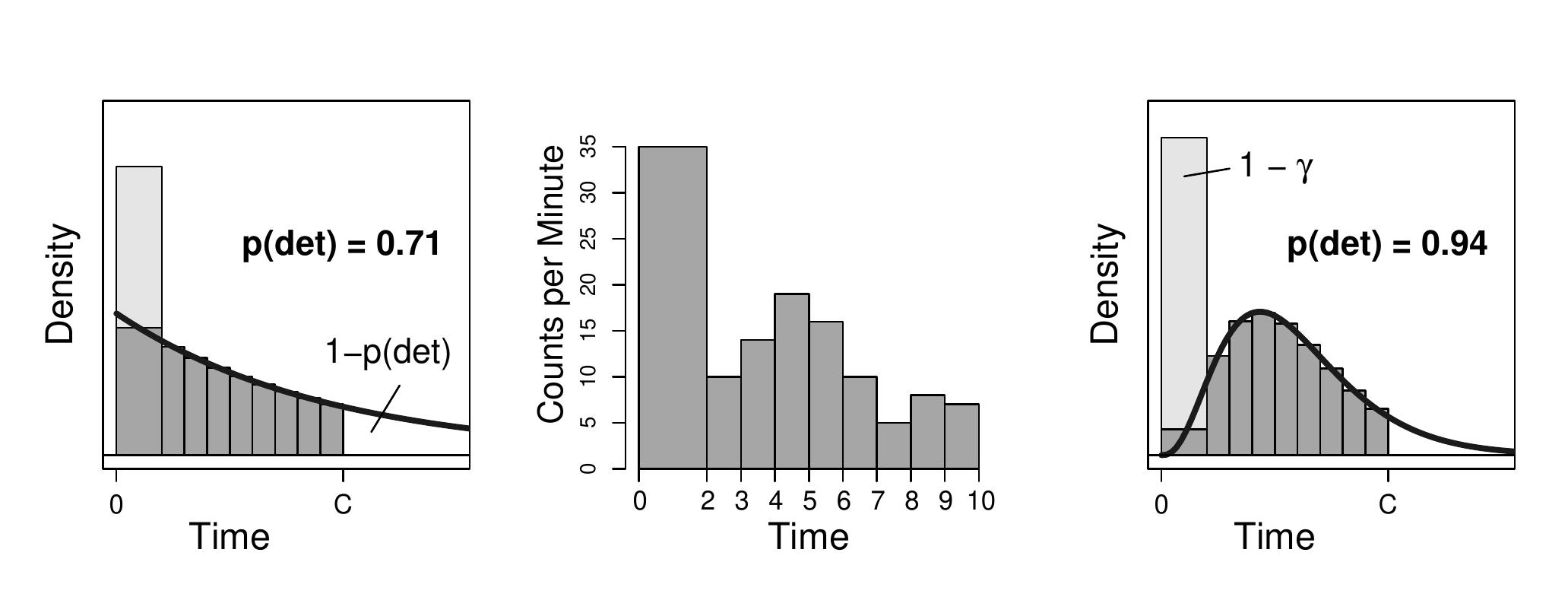}
\caption{\label{fig:schematic} 
Illustration of fitting mixture exponential (left) and mixture gamma (right) time-to-detection distributions (TTDDs) to interval-censored removal-sampled observations (center).
The mixture TTDD consists of a continuous TTDD (thick line) plus a mixture component of first-interval detections (light gray rectangle), constituting $\gamma$ and $1-\gamma$ proportions of the population, respectively.
We estimate $\pdet$ as the proportion of the TTDD before the end of the observation period $C$, leaving an estimated proportion $1-\pdet$ undetected.
}
\end{figure}

Framing a model in terms of time-to-detection leads to two practical differences vis-a-vis constant-detection models.  
First, in order to model covariate and random effects on detection, we perform mixed effects linear regression on the log of the rate parameter as in \citet{Solymos2013}, whereas most existing studies instead construct regression models on the logit of the equal-interval detection probability.  
The latter is not possible when detection rates are not constant.  
Second, because we can obtain interval-specific detection probabilities from the TTDD by partitioning its cdf (Figure \ref{fig:schematic}), we can directly model the data according to their existing interval structure rather than subdividing the observation period into intervals of equal duration.  
Indeed our model fits exact time-to-detection data, whereas existing constant-detection removal models only approximate exact data by subdividing the observation interval into a large number of fine equal-duration intervals \citep{Reidy2011, Amundson2014}.

Section \ref{sec:data} provides a description of the interval-censored time-to-detection avian point count data under consideration.
Section \ref{sec:model} introduces an N-mixture model with a generically defined TTDD for estimating abundance from removal-sampled point-count surveys.
Section \ref{sec:sim} provides three simulation studies to assess the impact of TTDD choice on estimated detection probability. 
Section \ref{sec:ovenbirds} analyzes an Ovenbird data set under different TTDDs to determine the impact of this choice on estimated detection probability and therefore estimated abundance.

\section{Interval-censored point counts}\label{sec:data}

Our analysis is motivated by avian point-count surveys in Chippewa National Forest from 2008-2013 as part of the Minnesota Forest Breeding Bird Project (MNFB) \citep{Hanowski1995}.  
For our analysis, we focus on Ovenbird counts selected from one habitat type: sawtimber red pine stands with no recent logging activity.  
Each stand had up to four sites with sufficient geographical distance between sites to reduce or eliminate overlapping territories.
This dataset includes 947 Ovenbirds counted in 381 surveys at 65 sites with site-specific variables including site age,  stock density, and an indicator of select-/partial-cut logging during the 1990s.

Single-visit (per year) point-count surveys were conducted by trained observers at each site once annually (weather permitting).  
Fourteen different observers conducted surveys during the study period and 69\% of surveys in our dataset involved observers in their first year at the MNFB.  
Survey durations were 10 minutes, with times to first detection censored into nine intervals: a two-minute interval followed by eight one-minute intervals.  
During each survey, the Julian date, time of day, and temperature were recorded. 

While we focus on the estimation of detection probability in avian populations, the approach we describe is appropriate for point-count surveys of any species.
The methodology allows the analysis of data with 1) recorded first (possible censored) detection of each individual, 2) site-specific explanatory variables, and 3) survey-specific explanatory variables.  

\section{Continuous time-to-detection N-mixture models} \label{sec:model}

Before considering interval censoring and explanatory variables, we first present the scenario of exact times to detections with no explanatory variables. 
We then incorporate interval censoring and follow with inclusion of fixed and random effects for abundance and detection. 

\subsection{Distributions for exact times to detection}\label{sec:exact_time}

Suppose that, for each survey $s$ ($s=1,\dots,S$), $N_s$ individuals are present.  
Imagine an observer could remain at the survey location until every individual is detected, recording the time to detection $t_{sb}$ (for bird, $b=1,\dots,N_s$) for each.
Assuming detection times for all individuals at a survey are independent, identically distributed according to a common time-to-detection distribution (TTDD), we define $T_{sb}$ as a random variable with cumulative distribution function (cdf) $F_T(t)$ and probability density function (pdf) $f_T(t)$.
In practice, times to first detection are truncated due to a finite survey length of $C$, meaning each individual has detection probability $\pdet=F_T(C)$.  
The conditional distribution of observed detection times consequently has pdf $\ftdt= f_T(t)/F_T(C)$ for $0<t<C$, cdf $F_{T|\dt}(t|\dt) = \int_0^t f_{T|\dt}(x|\dt) dx$, and instantaneous detection rate, or hazard function, $h(t) = f_T(t) / [1-F_T(t)]$. 

A common choice for TTDD is an exponential distribution, i.e. $T_{sb}\ind \mbox{Exp}(\varphi)$, which imposes a constant first detection rate, i.e. $h(t) = \varphi$.
By choosing another TTDD, we can allow for a systematic non-constant detection regime. 
For example, to model an observer effect where: (i) the observer's arrival suppresses or stimulates detectable cues, but (ii) individuals acclimate and gradually return to constant detection, a gamma TTDD would be appropriate.
Like the gamma TTDD, Weibull and lognormal TTDDs offer the flexibility of a two-parameter form and allow rates to increase or decrease during the survey.
All three TTDDs may provide reasonable empirical approximations of non-constant detection, though the shapes of the distributions differ, potentially leading to differing inference.
For instance, when detection rates vary across individuals, the result is a marginal detection rate that decreases over time.
Whether the marginal rate is best approximated by a gamma, lognormal, Weibull, or some other TTDD depends on just how rates vary across individuals.

To facilitate the later inclusion of fixed and random effects, we use the following rate-based parameterizations: $T\sim \Exp(\varphi), E[T]=1/\varphi$; $T\sim \Ga(\alpha,\varphi), E[T] = \alpha/\varphi$; $T\sim \We(\alpha,\varphi), E[T]=\mathrm{\Gamma}(1+1/\alpha)/\varphi$; and $T\sim \LN(\varphi,\alpha), E[T] = \exp(\alpha^2/2)/\varphi$.  
This parameterization of the lognormal relates to the standard ($\mu, \sigma^2$) parameterization by $\varphi = \exp(-\mu)$ and $\alpha = \sigma$.  
The exponential distribution is a special case of both the gamma and Weibull distributions when $\alpha=1$. 
We employ a log link to model $\varphi$, and therefore our model is equivalent to a generalized linear model with a log link on the mean detection time.

\subsection{TTDDs in an N-mixture model}

A basic N-mixture model describes observed survey-level abundance $\nobs$ with a hierarchy where $\nobs_s \ind \mbox{Binomial}\left(N_{s}, \pdet\right)$ and $N_s \ind Po(\lambda)$.
We can decompose $N_{s}$ into observed and unobserved portions: $\nobs \ind Po(\lambda \pdet)$ and, independently, $n_{s}^{(unobs)} \ind \Po\left(\lambda[1-\pdet]\right)$.
Although alternative distributions can be considered, e.g. negative binomial, our experience with Ovenbird point counts suggests that, after accounting for appropriate explanatory variables, the resulting abundances are likely underdispersed rather than overdispersed, and thus we will use the Poisson assumption here. 

The above definitions complete our exact-time homogenous-survey data model, consisting of distributions for counts and observed detection times:
\begin{align}
\nobs_s &\ind \Po(\lambda \pdet) \nonumber\\
p(\textbf{t}_s) &= \prod\limits_{b=1 \dotso\nobs_s} f_{T|\dt}(t_{sb} | \det, \alpha, \varphi) \label{eq:simplemodel}\\
\pdet &= F_T(C|\alpha,\varphi) \nonumber
\end{align}
where $\textbf{t}_s$ is a vector of observed times at survey $s$.

\subsection{Interval-censored times to detection} \label{s:interval}

Due to the harried process of avian point counts, times to first detection are typically not recorded exactly, but are instead censored into $I$ intervals. 
Let $C_i$ for $i=1,\dots,I$ indicate the right endpoint of the $i$th interval then $C_I$ is the total survey duration and, letting $C_0=0$, the $i$th interval is $(C_{i-1},C_{i}]$. 
Let $n_{si}$ be the number of individuals counted during interval $i$ on survey $s$, $\nobs_s = \sum_{i=1}^I n_{si}$, and $\vn_{s}=(n_{s1},\dots,n_{sI})$.
Assuming independence amongst individuals and sites, we have $\vn_{s} \ind \Mult \left(\nobs_s, \vp_{s}\right)$, where $\vp_{s}=(p_{s1},\dots,p_{sI})$ and $p_{si} = F_{T|\dt}(C_i|\dt) - F_{T|\dt}(C_{i-1}|\dt) = \int_{C_{i-1}}^{C_i} f_{T|\dt}(t) dt$, 
see Figure \ref{fig:schematic}.  

\subsection{Detection heterogeneity across subgroups} \label{s:subgroups}

It is common in avian point counts to observe increased detections in the first interval relative to an exponential distribution.
This is often understood to reflect unmodeled detection heterogeneity across behavioral groups in the study population.
Failure to account for such heterogeneity in the constant-detection scenario leads to negative bias in abundance estimates \citep{Otis1978}.
To accommodate this empirical observation, many models of interval-censored removal times define a TTDD with a mixture component to increase the probability of observing individuals in the first interval \citep{Farnsworth2002, Royle2004Generalized, Farnsworth2005, Alldredge2007, Etterson2009, Reidy2011}.
We specify a mixture TTDD with mixing parameter $\gamma\in[0,1]$, a point-mass during the first observation interval, and a continuous-time detection distribution $\Fm(t)$.
The mixture TTDD cdf is defined: $F_T(t) = (1-\gamma) + \gamma \Fm(t)$ for $t>0$.
If $\gamma=1$, the non-mixture model is recovered.

\subsection{Incorporating explanatory variables}\label{sec:covariates}

As discussed in Section \ref{sec:data}, explanatory variables are available for sites and for surveys. 
Generally, we suspect that site variables, e.g. habitat, will affect abundance and survey variables, e.g. time of day, will affect detection probability. 
Thus, we allow for incorporating explanatory variables on both the abundance and detection.

To incorporate explanatory variables on abundance, we model the expected survey abundance $\lambda_{s}$ with log-linear mixed effects, i.e. $\log (\lambda_{s}) = \vX_{s}^A\vbeta^A + \vZ_{s}^A\vxi^A$ where $\vX_{s}^A$ are explanatory variables, $\vbeta^A$ is a vector of fixed effects, $\vZ_{s}^A$ specifies random effect levels, and $\xi_j^A \ind N(0,\sigma_{A[j]}^2)$ are random effects where $A[j]$ assigns the appropriate variance for the $j$th abundance random effect.  

To incorporate explanatory variables on detection probability, we let the continuous portion of the TTDD depend on the explanatory variables through the now site-specific parameter $\varphi_s$. 
Specifically, we model $\log(\varphi_{s}) = \vX_{s}^D\vbeta^D + \vZ_{s}^D\vxi^D$, where $\vX_{s}^D$ are explanatory variables, $\vbeta^D$ is a vector of fixed effects, $\vZ_{s}^D$ specifies random effect levels, $\xi_j^D \ind N(0,\sigma_{D[j]}^2)$ are random effects where $D[j]$ assigns the appropriate variance for the $j$th detection random effect.  
For simplicity, we assume the shape parameter $\alpha$ as constant across sites.

\subsection{Estimation}

For ease of reference, the final full model is provided in equation \eqref{eq:model} where the conditioning of the TTDD cdf on $\alpha$ and $\varphi_s$ is made explicit.
\begin{align}
\nobs_s &\ind \Po(\lambda_s \pdet_s) \nonumber\\
\vn_s &\ind \Mult(\nobs_s, \vp_s); \qquad \vp_s = (p_{s1},\dots,p_{sI}) \nonumber\\
\pdet_s &= F_T(C_I|\alpha,\varphi_s) \nonumber \\
p_{s1} &= \left[(1-\gamma) + \gamma\Fm(C_1|\alpha,\varphi_s)\right]/\pdet_s \label{eq:model} \\
p_{si} &= \gamma\left[\Fm(C_i|\alpha,\varphi_s) - \Fm(C_{i-1}|\alpha,\varphi_s)\right]/\pdet_s \nonumber\\
\log(\lambda_s) &= \vX_{s}^A\vbeta^A + \vZ_{s}^A\vxi^A; \qquad \xi_j^A \ind N(0,\sigma_{A[j]}^2) \nonumber\\
\log(\varphi_{s}) &= \vX_{s}^D\vbeta^D + \vZ_{s}^D\vxi^D; \qquad \xi_j^D \ind N(0,\sigma_{D[j]}^2) \nonumber
\end{align}
We adopt a Bayesian approach and therefore require a prior over the model parameters.
To ease construction of a default prior for this model, we standardize all explanatory variables and then construct priors to be diffuse within a reasonable range of values.
Normal prior mean and standard deviation (sd) for the abundance intercept was set at a median abundance of 3 birds per site and a 95\% probability of 0-14 birds present (counted and uncounted).  
Normal prior mean and sd for the detection intercept were chosen so that, based on an intercept-only non-mixture model with $\alpha=1$: (i) median prior detection probability was $p_{s}^{(det)} = 0.50$, and (ii) 95\% of the prior detection probability was within $p_{s}^{(det)} \in (0.01, 1.0)$.  
Normal priors for fixed effect parameters were centered at zero with standard deviations matching the appropriate intercept term.  
All standard deviations and $\alpha$ were given half-Cauchy priors with location 0 and scale 1 for the untruncated Cauchy, and the mixture parameter $\gamma$ was assigned a Unif(0,1) prior in mixture models.
All scalar parameters were assumed independent \emph{a priori}.

We fit the models by MCMC sampling using the Bayesian statistical software Stan, implemented via the R package \texttt{rstan} version 2.8.0 \citep{Rstan2016}.  
We discarded half of the iterations as warmup and thinned by 10.  
We monitored convergence of the MCMC chains using Geweke z-score diagnostics \citep{Geweke1991} and reran models if lack of convergence was indicated by a non-normal distribution of the z-scores or if the effective sample size for any parameter was below 1000.  
The number of iterations used depended on the model and is detailed later. 
For most models, we accepted Stan defaults for initial values; however, gamma and Weibull models sometimes failed to run unless care was taken in the specification of initial values.

\section{Simulation studies} \label{sec:sim}

We conducted three simulation studies to explore the behavior of models with non-constant TTDDs.
The first study compares mixture vs non-mixture models.
The second study compares the TTDD families.
In the first two studies, we utilized intercept only models to focus attention on robustness of the TTDD choice in the most simple of scenarios.
For the third study, we included fixed and random effects for both abundance and detection and again compared the distribution families. 
In all simulation studies, we focused on accuracy in estimation of $\pdet$ which then translated into estimation of abundance. 

In the following analyses we distinguish two categories of purely continuous TTDDs: peaked and nonpeaked.  
Detection rates $h(t)$ of peaked distributions generally increase over time, while detection rates of nonpeaked distributions generally decrease over time.
More formally, we define a peaked TTDD as having a mode greater than zero (or $C_1$ for lognormal) while a nonpeaked TTDD has a mode of zero (or less than $C_1$), but we consider exponential TTDDs to be neither peaked nor nonpeaked.

\subsection{Mixture versus non-mixture TTDDs}\label{sec:mixture}

To assess the need for incorporating a mixture component to increase the probability of detection in the initial interval as discussed in Section \ref{s:interval}, we simulated 5600 intercept-only datasets: 100 replicates using 4 values of $\pdet$ (0.50, 0.65, 0.80, and 0.95) from each of 14 TTDDs (each combination of peaked/nonpeaked, mixture/non-mixture, and exponential/gamma/Weibull/lognormal, where exponential models are considered nonpeaked).  
We chose true parameter values (Table S-1) and the number of surveys (381) to mimic the Ovenbird analysis (Section \ref{sec:ovenbirds}).  
In particular, we set parameters such that (i) in nonpeaked models, 70\% of \textit{detected} individuals were observed during the first two minutes, and (ii) in peaked models, the detection mode for `hard to detect' individuals occured at 5 minutes.

We fit each dataset with two models: mixture and non-mixture versions of the distribution family, e.g. exponential, used to simulate the data.
For each dataset-analysis combination, we sampled $>$ 90,000 iterations which showed no evidence of lack of convergence according to the Geweke diagnostic and reached over 1,000 effective samples for all parameters. 

\begin{figure}\centering
\includegraphics[width=\textwidth]{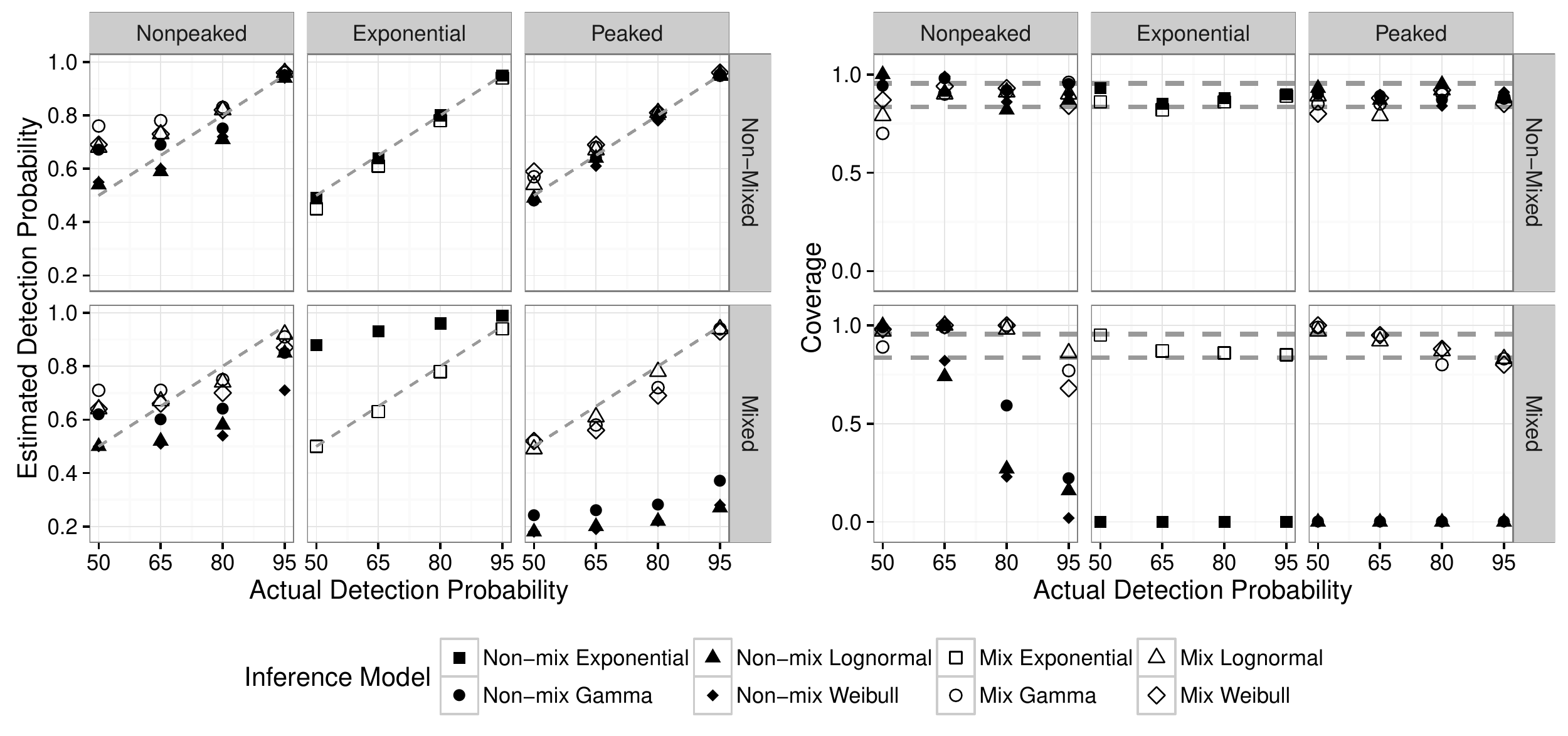}
\caption{
Comparisons of mixture and non-mixture TTDDs.  \textit{Left}: Average posterior median detection probabilities across 100 replicate simulations.  Dashed lines show unbiased estimation.   \textit{Right}: Coverage of 90\% credible intervals across 100 replicate simulations.  Dashed lines depict a 95\% range of observed coverage that is consistent with nominal coverage.  \textit{Rows}: Data simulated from non-mixture (upper) or mixture (lower) TTDDs.  \textit{Columns}:  Data simulated from nonpeaked (left), exponential (center), or peaked (right) TTDDs.  Each inference model was fit only to datasets from the same TTDD family (e.g. lognormal to lognormal).
}
\label{fig:mixture_fig} 
\end{figure}

For each dataset and inference model combination, we summarize the analysis across simulations by averaging the posterior median and reporting coverage for 90\% credible intervals.
If analyses are providing reasonable estimates of $\pdet$, we expect the average median to be unbiased and the coverage to be close to 90\%.
Figure \ref{fig:mixture_fig} provides a summary of these quantities. 
When a mixture model is used to simulate the data (lower row of plots), there is clearly a benefit to using a mixture model for inference.  
Using a non-mixture model for inference, the credible interval coverage is near zero for most models with the exponential model overestimating $\pdet$ and the other models underestimating. 
When a non-mixture model is used to simulate the data (upper row of plots), there are no clearly discernable differences between the ability of a non-mixture or mixture model to capture $\pdet$. 
These results support the general default use of a mixture model over a non-mixture model.

For nonpeaked datasets, estimates of $\pdet$ from the same TTDD inference model differed by only 1-5\% between $\pdet=0.50$ and $\pdet=0.65$ simulations (Tables S-3 to S-6), and credible intervals had roughly the same widths.
This suggests that, when data are nonpeaked with true detection probabilities less than 65-80\%, the patterns of detections over time are insufficient for distinguishing between moderate and low values of $\pdet$.
In these cases, mixture inference models estimated higher detection probabilities than did non-mixture models.

\subsection{Constant vs. non-constant detection mixture TTDDs}\label{sec:family}

The previous section addressed model mis-specification in terms of the mixture component. 
Now we turn to model misspecification of the distribution family. 
We simulated 100 replicates of intercept-only datasets from the 7 different mixture TTDD models using the same detection probabilties and parameters as in the previous section, and we fit them with mixture models from each of exponential, gamma, lognormal, and Weibull families.

\begin{figure}\centering
\includegraphics[width=\textwidth]{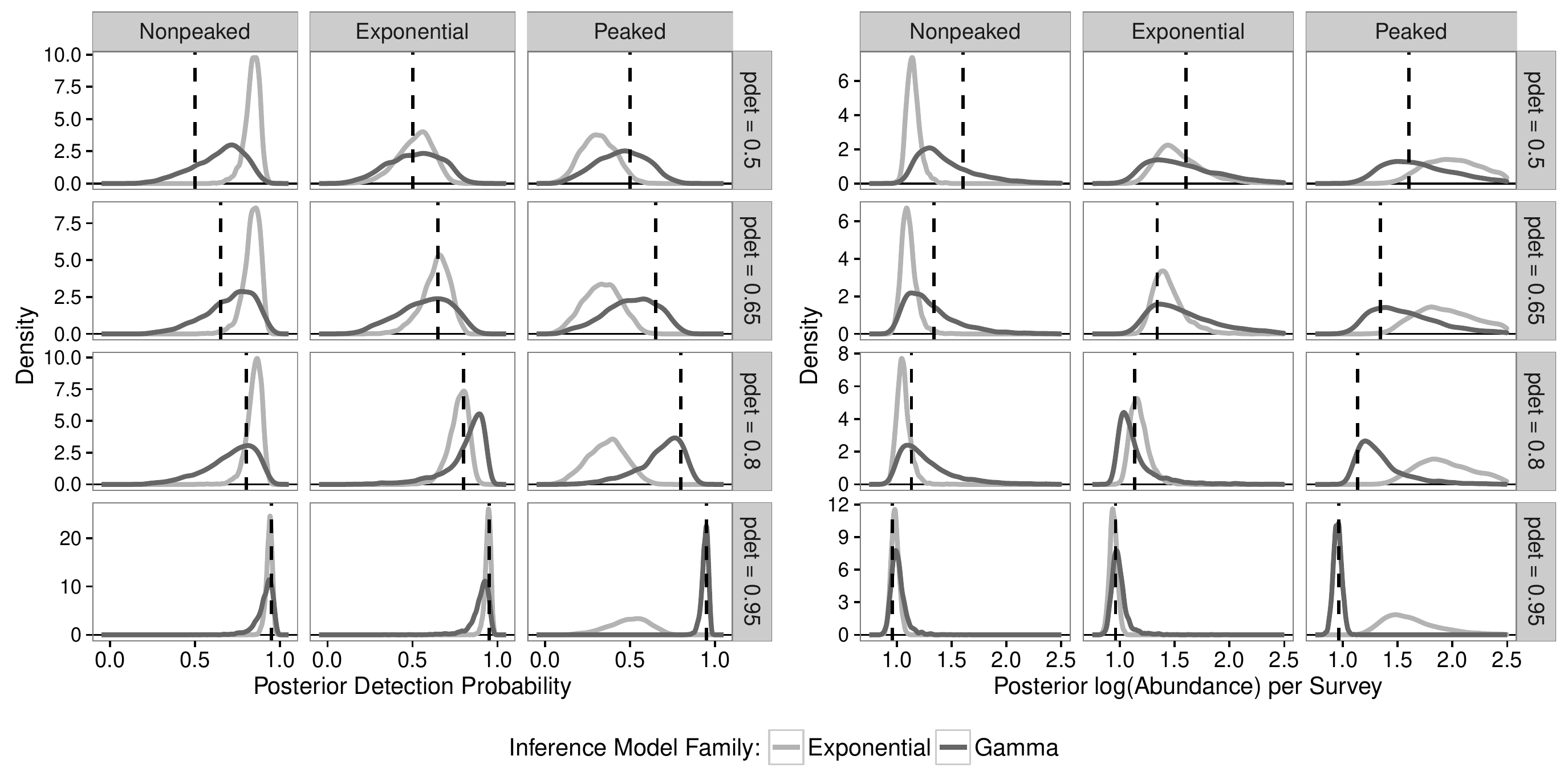}
\caption{\label{fig:sim2typical} 
Representative examples of posterior distributions for $\pdet$ (left panel) and survey-level log-scale abundance (right panel).
Data are simulated from nonpeaked gamma (left column), exponential (center), and peaked gamma (right) mixture TTDDs.
True detection probabilities vary by row and are shown for comparison (dashed vertical line).
Inference models are either exponential mixture (light gray) or gamma mixture (dark gray) TTDDs.
}
\end{figure}

Figure \ref{fig:sim2typical} illustrates representative examples of posterior distributions for $\pdet$ and site-level log(abundance) when data and models were from the exponential and gamma mixture TTDDs. 
Posterior distributions under exponential inference models accurately captured true detection probabilities when the simulation model was exponential, but they overestimated (underestimated) detection probabilities when the simulation model was nonpeaked (peaked).
In contrast, gamma family posteriors, which have added flexibility from having a shape parameter, accurately captured the truth in most scenarios although with increased uncertainty. 

\begin{figure}\centering
\includegraphics[width=\textwidth]{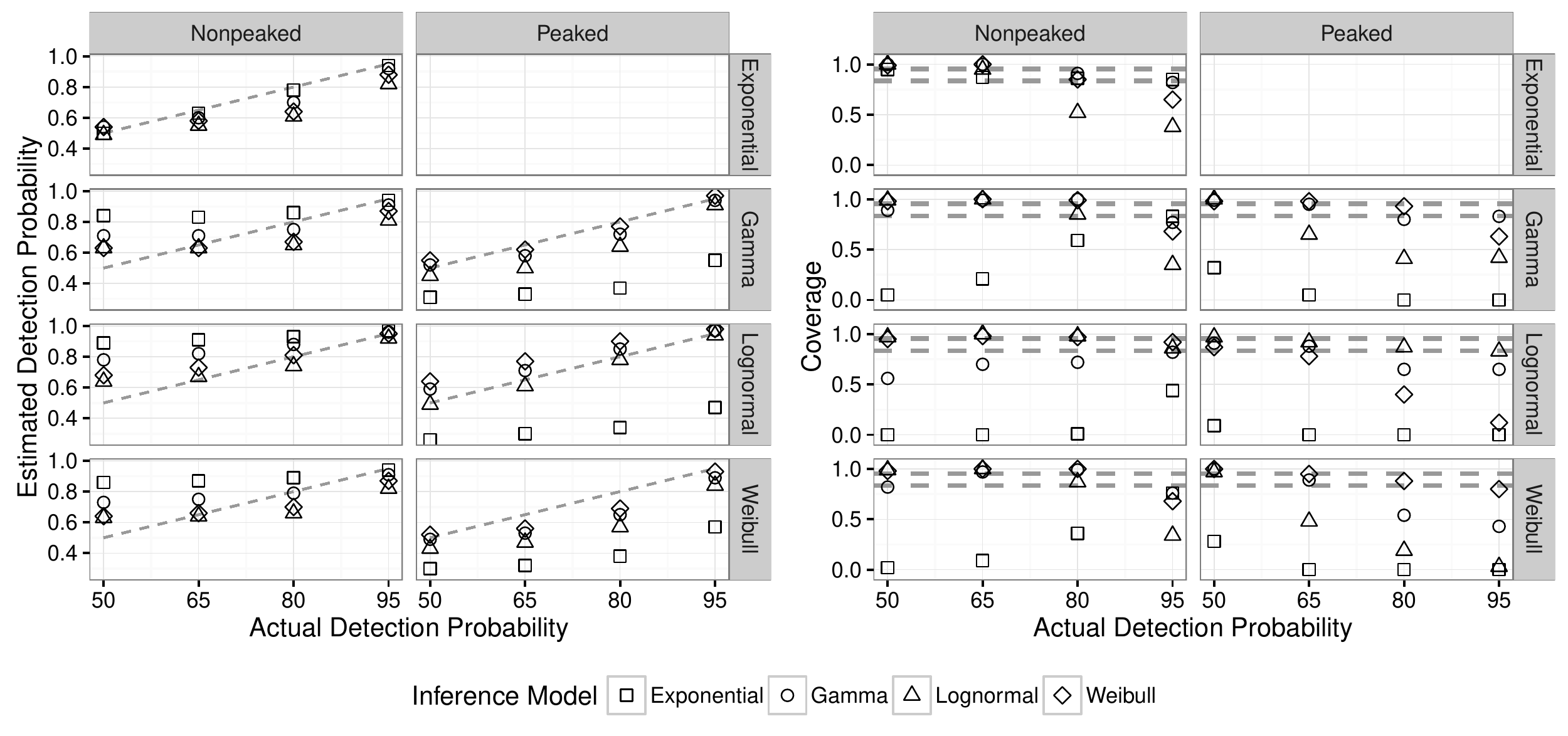}
\caption{
Comparisons of TTDDs across families.  \textit{Left}: Average posterior median detection probabilities across 100 replicate simulations.  Dashed lines show unbiased estimation.   \textit{Right}: Coverage of 90\% credible intervals across 100 replicate simulations.  Dashed lines depict a 95\% range of observed coverage that is consistent with nominal coverage.  \textit{Rows}:  Data simulated from mixture exponential, gamma, lognormal, or Weibull TTDDs.  \textit{Columns}: Data simulated from nonpeaked or exponential (left) or peaked (right) TTDDs.  All data and inference models used mixture TTDDs.
}
\label{fig:family_fig}
\end{figure}

Figure \ref{fig:family_fig} and Tables S-7 to S-10 summarize posterior estimates of $\pdet$ and coverage from 90\% credible intervals across all TTDD families.
The poorest estimation of $\pdet$ occurred for the exponential inferential model when the data simulation model had a peak, because the two parameters (rate and mixing parameter) did not provide enough flexibility to adequately fit a TTDD with both an initial increase and a delayed mode. 
As a result, the exponential model underestimated the actual detection probability.
In contrast, the exponential model typically overestimated detection probability for nonpeaked simulated data.
However, exponential model estimates were both less biased and more precise when the data actually derived from an exponential mechanism.

When comparing the different three-parameter TTDDs, model misspecification was not as serious an issue because the models could better account for the patterns in time to detection.
Even so, estimates of $\pdet$ amongst the three models differed by as much as 0.15.
As in the previous simulation study, estimates of $\pdet$ from nonpeaked datasets changed little as true values of $\pdet$ decreased below 65-80\%.
For nonpeaked datasets, gamma TTDDs produced larger estimates of $\pdet$ than did Weibull TTDDs, with lognormal TTDDs producing the lowest of all.
For peaked datasets, the order of Weibull estimates were larger than gamma estimates.
Our results do not favor the use of one of these TTDDs over the others.

\subsection{Models including covariates and random effects}\label{sec:simfull}

The previous sections studied effects of time to detection assumptions in the context of no explanatory variables.
We now incorporate fixed and random effects for abundance and detection to ascertain whether models differ in their estimates of effect sizes.
We simulated data from each of the 7 mixture TTDDs and fit models from exponential, gamma, lognormal, and Weibull mixture models.  

We simulated data using the median posterior parameter estimates obtained in the analysis of Ovenbird data in Section \ref{sec:ovenbirds}.
As those models differed in their estimates of $\pdet$, so the simulated datasets featured different true values of $\pdet$.
Because fitted Ovenbird models did not yield peaked distributions, we simulated peaked data by: (i) using the same intercepts, shape parameters, and mixing parameters as for peaked data ($\pdet = 0.80$) in previous simulations, (ii) using median covariate and random effects from the Ovenbird estimates, and (iii) scaling the detection intercept and random effect to achieve true detection probabilities $\approx 0.8$ with a detection mode at 5 minutes.
See Table S-2 for actual parameter values.
Because of the difficulty in integrating random effects over all sites, approximate posterior distributions for the study-wide marginal $\pdet$ were obtained by simulating data from each MCMC sample and calculating the proportion of simulated Ovenbirds that were observed.

Computation times for this simulation study were much greater than for the other studies because partitions of the cdf, e.g. $\Fm(C_i|\alpha,\varphi_s)$, had to be calculated separately for every survey; also, sampling often required 2-8 as many iterations.
Average times for exponential, lognormal, Weibull, and gamma model fits were 2.6, 5.1, 5.3, and 25+ hours, respectively, as compared to only 2.0, 3.0, 3.3, and 5.3 minutes for the intercept-only models.
Due to the computation times involved, we fit each model only once.

The results from this simulation are qualitatively similar to the Ovenbird analysis (Figure \ref{ovenposteriors}) and thus we only briefly review the results here and provide the corresponding figures and tables in the Supplementary Material. 
Patterns in posterior estimates of overall detection probabilities with respect to family TTDD forms were largely the same as in the previous simulation studies for appropriate values of $\pdet$ -- the inclusion of explanatory variables did not make models more robust to violations of constant-detection assumptions (Figure S-1).  
Posteriors for abundance-related fixed and random effects were the same regardless of which TTDD was assumed (Section S-2.1 and figures therein).  
Posteriors for the mixing parameter $\gamma$ and detection-related fixed and random effects were the same across gamma, lognormal, and Weibull mixture models but were narrower and location-shifted for the exponential mixture model.  

\section{Ovenbird analysis}\label{sec:ovenbirds}

We fit the Ovenbird dataset with exponential, gamma, lognormal, and Weibull mixture models.  
For the abundance half of our model, we used four covariates plus two random effects.  
The covariates were: (a) site age, (b) survey year, (c) an indicator of whether the site stock density was over 70\%, and (d) an indicator of whether the site experienced select-/partial-cut logging during the 1990s.  
We associated random effects with each survey year and each stand.
For the detection half of our model, we used covariates for: (a) Julian date, (b) time of day, (c) temperature, (d) an indicator of whether it is the observer's first year in the database, and (e) an interaction between (a) and (d) to approximate a new observer's learning curve.  
We associated random effects with each observer.  
Preliminary model fits did not support the inclusion of quadratic terms for any detection covariates.  
We centered and standardized all continuous covariates prior to fitting models.
We ran chains 250,000-375,000 iterations; Geweke diagnostics showed no indication of lack of fit, and effective sample sizes were over 1000 for all parameters.

\begin{figure}[h!]\centering
\includegraphics[width=0.95\textwidth]{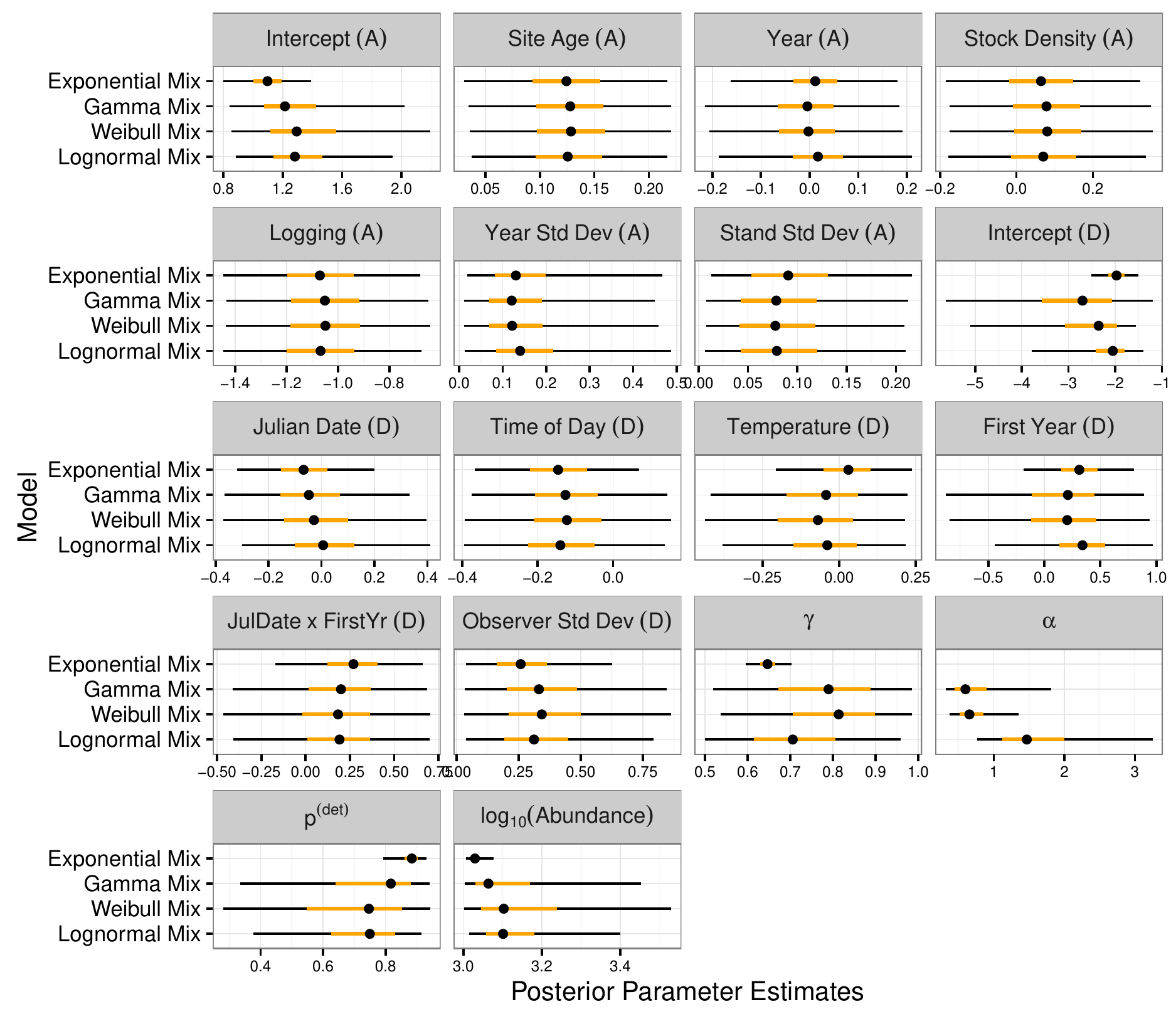}
\caption{\label{ovenposteriors} Posterior medians (black dots) with 50\% (wider line) and 95\% (narrowerline) credible intervals for the mixing parameter ($\gamma$), shape parameter ($\alpha$) as well as abundance (A) and detection (D) fixed effects and random effect standard deviations.
Posteriors are also available for the overall probability of detection and the abundance across all sites.}
\end{figure}

Figure \ref{ovenposteriors} presents posterior medians and credible intervals for model parameters, overall detection probability $\pdet$, and the logarithm of total Ovenbird abundance.  
Estimates for the shape parameter $\alpha$ from the gamma and Weibull models are consistent with the data arising from an exponential distribution, although the uncertainty on this parameter remains relatively large. 

Abundance covariate coefficient estimates were virtually the same across all models.  
The 95\% credible intervals for two of the abundance parameters (site age and logging) do not contain zero, thereby suggesting notable effects.  
Select- and partial-cut logging events of the 1990s depressed local Ovenbird abundance during the study perior to roughly 25-50\% of the abundance for unlogged sites.  
Credible intervals for site age coefficient indicate that each decade of age increases abundance from 1.5-13\%.
Credible intervals for detection parameters do not indicate significant effects, after adjusting for the other predictors, for any of the included predictors.

In spite of the similarity of effect parameter estimates, the posterior distributions for detection probability and abundance differ greatly between the exponential and non-exponential models.  
It is clear that the assumption of constant detection leads to much higher and more precise estimates of detection than would be obtained if we are unwilling to make that assumption.

\section{Discussion} \label{sec:discuss}

We formulated a model for single-species removal-sampled point-count survey data that allows for non-constant detection rates.
The model accommodates both interval-censored and exact times to detection.
Our model adopts a time-to-event approach within a hierarchical N-mixture framework, and it allows times to first detection to be modeled according to flexibly defined TTDD families.
Our results show that non-constant TTDDs can return reasonable estimates of detection probabilities across a variety of time-to-detection data patterns, whereas traditional constant-rate TTDDs return biased and overly precise estimates when data deviate from the constant-rate assumption, even when they include a mixture for heterogeneity across groups.  
Because the exponential TTDD is a special case of both gamma and Weibull TTDDs, we can interpret the differences in estimation between models as resulting from the information conveyed by the assumption of constant detection.

We have additionally demonstrated for non-constant models the utility of using a mixture TTDD formulation.
Inference models with a mixture component are accurate under most scenarios whether the data have a mixture or not, whereas inference models without the mixture can be badly biased when the data do feature a mixture.  

If the estimation of effect parameters and the roles of explanatory variables are the primary interest, then our results suggest that the exact choice of TTDD may not be important.  
Abundance effect estimates are similar regardless of the chosen TTDD.
Detection effect estimates, while conditional on the mixing parameter $\gamma$, are similar across all mixture non-exponential TTDDs.  
These findings may well not hold if the same covariate is modeled in both abundance and detection models \citep{Kery2008}.

If the estimation of abundance is the primary interest, then the choice of TTDD has large consequences, and we may reasonably ask when removal sampled point-count surveys are adequate.
Based on our results, we would be skeptical of abundance estimates that indicate a nonpeaked TTDD with median posterior estimates of $\pdet$ below 0.75 (using a mixture Weibull as a yardstick).
Reduced model performance in the presence of low detection probabilities is common among abundance models.
We recommend the discussion in \citet{CoullAgresti1999}, elaborating key points here.
The essential problem is a flat log-likelihood.
For single-pass surveys, we cannot rely on repeated counts to improve our estimates.
Instead, the strategy of removal sampling is to use the observed pattern of detection times to estimate the proportion of individuals that would be detected if only the observation period lasted longer.  
As such, it is entirely reliant upon extrapolation based on an accurate fit of $\ftdt$.
When true detection probabilities are low, $\ftdt$ for nonpeaked data becomes flat.
Such a flat observed pattern provides little information and is well approximated by a wide variety of TTDDs, which vary greatly in their tail probabilities.
An assumption of constant detection limits the shape of $\ftdt$ and thereby constrains uncertainty but at the cost of potentially sizable bias.

Just as the exponential TTDD is a special case of the two-parameter gamma and Weibull TTDDs, so all four TTDDs in our analysis are special cases the three-parameter generalized gamma distribution.
A generalized gamma TTDD encompasses a diversity of hazard functions \citep{Cox2007}, 
eliminating the need to restrict analysis to lognormal, gamma, or Weibull TTDDs and the tail probabilities they imply.
However, maximum likelihood estimation of the generalized gamma has historically suffered from computational difficulties, unsatisifactory asymptotic normality at large sample sizes, and non-unique roots to the likelihood equation \citep{CoorayAnand2008, NoufailyJones2013}.
It may well be possible to implement our model with a generalized gamma TTDD, but especially when we consider the right-truncation of data from point-count surveys, we think model convergence would not be a trivial problem.

Study design can address some issues associated with non-constant detection.
In theory, longer surveys can improve estimation of the unsampled tail probability, but the longer the survey lasts, the greater the risk that individuals enter/depart the study area or are double-counted, which violates the removal sampling assumption of a closed population \citep{LeeMarsden2008, Reidy2011}.  
Observer effects on availability rates can be mitigated by introducing a settling down period but at the potential cost of a serious reduction in total observations \citep{LeeMarsden2008}.
An alternative to removal sampling is to record complete detection records (all detections for every individual) instead of just the first \citep{Alldredge2007}; however, this may not be feasible in studies like MNFB where many species are observed simultaneously

Versions of time-varying models have been described for trap-based removal sampling and continuous-time capture-recapture.  
Time variation has been modeled through a non-constant hazard function \citep{Schnute1983, HwangChao2002}, a randomly varying detection probability across trapping sessions \citep{WangLoneragan1996}, and constant detection probabilities that vary randomly from individual to individual \citep{Mantyniemi2005, Laplanche2010}.
Most of these approaches resulted marginally in a decreasing (nonpeaked) detection function over time.  
Their results generally echo what we have presented here.  
\citet{Schnute1983} found that the equivalent of a mixture exponential adequately described their data.
\citet{WangLoneragan1996}, \citet{HwangChao2002}, and \citet{Mantyniemi2005} all found constant-detection models to be flawed, producing underestimates of abundance and too-narrow error estimates; these resulted in inadequate coverage and also overstatement of effect significance.  

Point-count survey data often include the recorded distance between observer and detected organism.
Because our focus has been on modeling variations in detection rates during the survey period, we have not incorporated distance into our model.
Consequently, our application of a TTDD represents an averaging across distance classes, which induces systematic bias in estimates of abundance \citep{EffordDawson2009, Laake2011, Solymos2013}.  
To be consistent with the continuous time-to-event approach, distance can be incorporated into the detection model as an event-level modifier as is done in \cite{Borchers2016}.  
This approach is distinct from earlier integrations of removal- and distance sampling, where distance has been treated as an interval-/survey-level modifier \citep{Farnsworth2005, Diefenbach2007, Solymos2013, Amundson2014}.
The differences between these implementations may impact estimates of detection and abundance, especially in the presence of behavioral heterogeneity in availability rates across subgroups of the study population.  
This is an area of ongoing exploration.

We recommend that time-heterogeneous detection rates be explicitly modeled in single-species analyses involving removal-sampled point-count survey data where estimation of detection probability or abundance is a primary objective.  
The assumption of constant detection, while computationally simple and reasonable as a null model, proves to be rather informative and can result in pronounced bias.  
Meanwhile, the causes of non-constant detection -- i.e., observer effects on behavior and systematic variations in observer effort -- are both plausible and not trivially discounted.  
It would be nice if the data itself could inform us whether constant detection is a reasonable assumption; however, our preliminary efforts to diagnose this assumption using deviance information criterion (DIC) and posterior predictive check statistics have led to weak and sometimes erroneous findings.
Development of such a diagnostic tool would be useful, but given the limitations of first time-to-detection data, we are not confident a reliable tool could be easily developed.  
We believe that more informative data collection, such as complete time-to-detection histories and microphone arrays, offer more effective tools for time-to-event modeling going forward.

\section{Supplementary Materials}

The supplementary materials include supplementary tables and figures.

\bibliography{masterbib}
\bibliographystyle{biom}

\end{document}